\documentclass[aps,prl,superscriptaddress,twocolumn,preprintnumbers]{revtex4}

\usepackage{amsmath,amsfonts,amssymb,amsthm,amstext,amscd,eucal}
\usepackage{graphicx}

\usepackage{color}

\newcommand{\bR}{\mathbb{R}}

\def\beq{\begin{equation}}
\def\eeq{\end{equation}}
\def\beqa{\begin{eqnarray}}
\def\eeqa{\end{eqnarray}}

\def\IS{{\mathbb{S}}}
\def\IR{{\mathbb{R}}}
\def\IZ{{\mathbb{Z}}}
\def\IX{{\IX}}

\def\tr{{\rm tr \,}}

\def\aDnine{{\overline{{\rm D}9}}}

\newcommand{\drawsquare}[2]{\hbox{%
\rule{#2pt}{#1pt}\hskip-#2pt
\rule{#1pt}{#2pt}\hskip-#1pt
\rule[#1pt]{#1pt}{#2pt}}\rule[#1pt]{#2pt}{#2pt}\hskip-#2pt
\rule{#2pt}{#1pt}}

\newcommand{\fund}{\raisebox{-.5pt}{\drawsquare{6.5}{0.4}}}
\newcommand{\Ysymm}{\raisebox{-.5pt}{\drawsquare{6.5}{0.4}}\hskip-0.4pt%
        \raisebox{-.5pt}{\drawsquare{6.5}{0.4}}}
\newcommand{\Yasymm}{\raisebox{-3.5pt}{\drawsquare{6.5}{0.4}}\hskip-6.9pt%
        \raisebox{3pt}{\drawsquare{6.5}{0.4}}}

\begin{document}
\preprint{MPP-2014-171}
\preprint{IFT-UAM/CSIC-14-035}
\preprint{FTUAM-14-15}

\title{Heterotic NS5-branes from closed string tachyon condensation}

\author{I\~naki Garcia-Etxebarria}
\email{inaki@mpp.mpg.de}
\affiliation{Max Planck Institute for Physics, 80805 Munich, Germany}

\author{Miguel Montero}
\email{miguel.montero@uam.es}
\affiliation{Departamento de F\'isica Te\'orica, Universidad Aut\'onoma de Madrid, 28049 Madrid, Spain}
\affiliation{Instituto de F\'isica Te\'orica UAM/CSIC, 28049 Madrid, Spain}

\author{Angel Uranga}
\email{angel.uranga@uam.es}
\affiliation{Instituto de F\'isica Te\'orica UAM/CSIC, 28049 Madrid, Spain}


\begin{abstract}
\noindent
We show how to construct the familiar heterotic NS5 brane as a
topological soliton in a supercritical version of heterotic string
theory. Closed string tachyon condensation removes the extra
dimensions, leaving the NS5 in 10d, in a process highly reminiscent of
the K-theoretical description of type II D-branes, but linking
non-trivial gauge bundles and geometry. The construction requires a
modification of the anomalous Bianchi identity for $H_3$ in
supercritical heterotic string theory. We give various proofs for the
existence of this modification.
\end{abstract}


\maketitle




\section{Introduction}

One of the main breakthroughs in string theory is the realization that it is not a theory of strings. Indeed, the theory includes a plethora of other extended objects, the branes, essential to define the non-perturbative regime, in which they are on equal footing with the fundamental string.  The best known such objects are D-branes, which admit a perturbative description in terms of open string sectors, and which underlie several deep connections, like the microstate interpretation of the Bekenstein-Hawking entropy for certain black holes and the AdS/CFT or gauge/gravity correspondence. An additional handle on D-branes is their description as solitons of open string tachyons, in extensions of the theory including brane-antibrane pairs, and their classification by K-theory \cite{Witten:1998cd}.   

Most other topological defects in string theory do not have a simple perturbative description or other tractable tools to construct them, except as supergravity solutions. A historically paradigmatic example is the heterotic NS5-brane, whose near-core region is inherently non-perturbative due to a blowing-up coupling, even if the asymptotic coupling constant is kept perturbative \cite{wshetinsts}. In this paper we show that the NS5-brane admits a description as a soliton of closed string tachyons in supercritical extensions of the theory. This opens up a new avenue to study these and possibly other objects in string theory. The construction is highly reminiscent of K-theory, implementing a physical equivalence of gauge field theory instantons and gravitational instantons, which may find application in other fields.

\section{Supercritical heterotic strings and closed string tachyon condensation}
\label{sec:heterotic-tachyon-cond}

Supercritical string theories \cite{Chamseddine:1991qu} can be
regarded as an extension of ordinary string theories by additional
dimensions, which can be removed by condensation of a closed string
tachyon in the spectrum \cite{Hellerman:2004zm,Hellerman:2006ff}, but
which can usefully display properties not manifest in the ordinary
critical theory \cite{Berasaluce-Gonzalez:2013sna}. In this work we
show that the $(10+n)$ dimensional supercritical extension $HO^{+(n)
  /}$ of the 10d $SO(32)$ heterotic string \cite{Hellerman:2004zm}
allows for a realization of heterotic branes as closed tachyon
solitons.

The worldsheet theory contains $D=10+n$ embedding coordinates
$X^M=X_L^M+X_R^M$ and $D$ right-moving fermions $\psi^M$, all in the
vector representation of the $SO(1,D-1)$ spacetime Lorentz group,
$M=0,\ldots, D-1$, and $32+n$ left-moving fermions $\lambda^a$,
$a=1,\ldots, 32+n$.
There is also a linear dilaton background to
achieve the central charge $c_{\rm matter}=15$, canceled against the
superghost sector. Modular invariance is achieved with a GSO
projection by $(-1)^F$,  anticommuting with $\psi^M$ and $\lambda^a$ (i.e. $F$ is the overall fermion number). The Neveu-Schwarz and Ramond
sectors correspond to the untwisted and twisted sectors with respect
to this $(-1)^F$ orbifold. The Ramond sector contains only massive
fields, so the light spectrum arises form the NS sector. It contains a
graviton $G_{MN}$, 2-index antisymmetric tensor $B_{MN}$, a dilaton
$\Phi$, and massless gauge bosons of $SO(32+n)$. In addition, there is
a tachyon $T^a$ in the vector representation of $SO(32+n)$, with
$\alpha'M^2=-2$.

This theory does not contain spacetime fermions. In order to include them and obtain a supercritical extension of the 10d supersymmetric  $SO(32)$  heterotic string, one performs a $\IZ_2$ orbifold, generated by an element $\theta$ acting as $\lambda^a\to -\lambda^a$, and $Y^m\equiv X^m\to -Y^m$, $\psi^m\to -\psi^m$, leaving $X^\mu$, $\psi^\mu$ invariant. Here we have split the indices as $\mu=0,\ldots, 9$, $m=10,\ldots, D-1$, so the orbifold breaks the Lorentz symmetry down to $SO(1,9)\times SO(n)$. 

The effect of $\theta$ in the parent theory is to project onto invariant states, in particular forcing the tachyons $T^a$, and the mixed tensors $G_{m\mu}, B_{m\mu}$, to vanish at the 10d fixed locus $X^m=0$. In addition, there are $\theta$-twisted sectors, localized at the 10d fixed locus, and which produce massless states in the sector where $\psi^m$ and $\lambda^a$ are antiperiodic. There, the groundstates are 10d spinors due to the fermion zero modes of $\psi^\mu$, whose chirality is fixed by the GSO projection. The massless fermions in this sector and their representations under the $SO(32+n)$ gauge group and the $SO(n)_{\rm rot}$ rotational group in the coordinates $X^m$ are as follows
\beqa
& \alpha_{-1}^\mu |{\rm spinor} \rangle \;\;& \rightarrow\quad   {\rm Gravitino+dilatino}\nonumber \\
&\lambda_{-\frac 12}^a\lambda_{-\frac 12}^b|{\rm spinor}_+\rangle\; & \rightarrow \quad  {\rm Fermion}_+\; {\rm in}\; (\Yasymm, 1) \nonumber \\
&\lambda_{-\frac 12}^a \alpha_{-\frac 12}^m|{\rm spinor}_-\rangle\; & \rightarrow\quad  {\rm Fermion}_-\; {\rm in}\;(\fund,\fund) \label{fermions} \\
&\alpha_{-\frac 12}^m\alpha_{-\frac 12}^n|{\rm spinor}_+\rangle\; & \rightarrow\quad    {\rm Fermion}_+\; {\rm in}\; (1,\Ysymm)\nonumber
\eeqa
where the $\pm$ subindex denotes the 10d spinor chirality.

This theory flows to the critical 10d string theory by condensation of the tachyon \cite{Hellerman:2006ff}, as follows. The tachyon profile $T^a(X,Y)$ couples in the worldsheet as a superpotential,
\beqa
\Delta {\cal {L}}=\frac 1{2\pi}\int d\theta_+\sum_a \lambda^a T^a
\eeqa
Splitting into component fields, and integrating out the auxiliary field, the tachyon Lagrangian is
\beqa
\Delta{\cal {L}}= -\frac 1{8\pi} \sum_a (T^a)^2\, +\,\frac i{2\pi} \sqrt{\frac{\alpha'}2}\,\partial_M T^a \, \lambda^a\psi^M
\label{tach-coupling}
\eeqa
Consider introducing a tachyon profile
\beqa
T^a\sim M^a_{\,n} Y^n
\label{het-tach}
\eeqa
where $M$ is some (momentarily constant) matrix. Notice that this ansatz is compatible with the  odd parity of the tachyon under the $\IZ_2$ orbifold, equivalently with the fact that the $\IZ_2$ symmetry is an R-symmetry of the 2d susy, under which the worldsheet superpotential must be odd. This tachyon deformation can be made exactly solvable by dressing (\ref{het-tach}) with a light-like exponential of $X^+$ \cite{Hellerman:2006ff}, but the key physics is already clear without it. For non-degenerate $M$, the potential term and fermionic couplings in (\ref{tach-coupling}) makes the supercritical coordinates massive, so the corresponding dimensions disappear upon tachyon condensation. The endpoint is critical 10d heterotic string theory, with the restriction of $\theta$ implementing the independence of the GSO projections for the left- and right-moving fermions. Lower rank matrices $M$ describe partial removal of the extra dimensions, effectively changing their number $n$. The analogy of the quantum numbers of fermions and tachyons in this theory and in type I with $n$ D9-$\aDnine$ brane pairs \cite{Sugimoto:1999tx} motivated a duality proposal between the two theories \cite{Hellerman:2004zm}. Our techniques and results do not rely on this analogy, and involve purely heterotic arguments; still, it is interesting that they agree with such matching, in particular relate the open and closed tachyons solitons and their K-theory constructions.

\section{Heterotic NS5-brane  \protect\\* from   tachyon condensation}

In this work we generalize the analysis to $X$-dependent tachyon background matrices, with rank jumping at particular loci. This arises for non-trivial bundles $V$, $V'$ for the $SO(n)\subset SO(32+n)$ gauge group and the geometric $SO(n)_{\rm rot}$ rotation group, so that the matrix $M=\partial T$ is a section of  $V\otimes V'{}^*$ and has zeroes over particular loci in the 10d slice. Tachyon condensation leads to the appearance of topological brane defects with worldvolume along those loci, allowing for the description of heterotic branes as closed string tachyon solitons. A particularly interesting case are heterotic NS5-branes, which are associated to non-trivial instanton backgrounds in these bundles. Actually, the relevant topological class is the first Pontryagin class of the {\em difference} bundle (in the K-theory sense). Quantitatively, the Poincar\'e dual 4-form to the NS5-brane worldvolume is
\beqa
\delta_4({\rm NS5})=\tr F^2_{\rm gauge}-\tr R^2_{\rm rot}
\label{ns-charge}
\eeqa The tachyon profile can be specified by the Atiyah-Bott-Shapiro
construction \cite{refabs}, whose local description is as
follows. Take a 4d subspace of the 10d slice, parametrized by
coordinates $\vec{X}$, in the theory with $n=4$ extra dimensions, and
regard the $SO(4)$ bundles $V$, $V'$ as chiral spinor bundles
$S^{\pm}$. Take the (derivative of the) tachyon as the map $S^+\to
S^-$ (determined by Dirac matrices $\Gamma$) \beqa
\partial_m T^a\sim \vec{\Gamma}_{ma}\cdot \vec{X}
\label{abs}
\eeqa
This tachyon soliton carries the appropriate topological charge, and upon tachyon condensation leaves a codimension 4 defect localized at the vanishing locus $\vec{X}=0$.

In order to show that this defect is actually an NS5-brane, the difference of instanton numbers in the bundles must be a magnetic source for the 2-form $B_2$, i.e. it must obey an anomalous Bianchi identity 
\beqa
dH_3=\tr F_{SO(32+n)}^2 -\tr R^2 -\tr R^2_{\rm rot}
\label{bianchi-new}
\eeqa
The first two terms are a generalization of the coupling in the
critical 10d heterotic theory; the extension to $SO(32+n)$,
required by gauge invariance and already noted in
\cite{Hellerman:2004zm}, reproduces the first term in
(\ref{ns-charge}). The second term is expected from invariance under
diffeomorphisms mixing the 10d slice and the supercritical dimensions
in the theory before the orbifold projection. Given its importance and
novelty, we give direct proofs using several arguments based on
anomaly cancellation in different dimensions. In addition, we also
give a worldsheet argument that the physics at $\vec{X}=0$ indeed
corresponds to the NS5-brane singular CFT \cite{wshetinsts}.

\subsection{Anomalies in 2d worldsheet}

The Bianchi identity (\ref{bianchi-new}) is in fact essential in the cancellation of anomalies of the worldsheet theory under target space diffeomorphisms and gauge transformations, as we now show. Using the worldsheet content, the relevant anomaly polynomial is
\beqa
I_4=\tr F_{SO(32+n)}^2 -\tr R^2 -\tr R^2_{\rm rot}\equiv dQ_3
\label{anomaly-pol-2d}
\eeqa
where we have introduced the gravitational, normal bundle and gauge Chern-Simons 3-forms (defined locally by $\tr F^2=d\omega_{\rm YM}$, etc), via 
\beqa
Q_3\equiv \omega_{3\, {\rm YM}}- \omega_{3\,{\rm L}}-\omega_{3\,{\rm rot}},
\eeqa
The 2d anomaly $\delta_\Lambda S_{2d}\equiv \Lambda I_2$ is obtained from the descent relations $\delta_\Lambda Q_3=\Lambda dI_2$, with $\Lambda$ a transformation parameter. This is canceled by an anomaly inflow as follows. The Bianchi identity (\ref{bianchi-new}) implies that
\beqa
H_3= dB_2 + Q_3
\eeqa
with the same Chern-Simons form $Q_3$.
The kinetic term produces interactions (in conventions of \cite{Polchinski})
\beqa
S_{B_2}=\int_{10+n } |H_3|^3\rightarrow \int_{10+n} *H_3 \wedge Q_3
\label{b-six-gs}
\eeqa
Since the string worldsheet $\Sigma_2$ is electrically charged under $B_2$, its Poincar\'e dual $\delta(\Sigma)$ acts as a source
\beqa
d*H_3=\delta(\Sigma_2)
\label{source-f1}
\eeqa
Target diffeomorphisms and gauge transformations produce a variation of (\ref{b-six-gs}) given by
\beqa
  \int_{10+n} *H_3 \wedge \delta_\Lambda Q_3= \Lambda\int_{10+n} *H_3 \wedge  dI_2= -\int_{\Sigma_2} I_2\nonumber
\eeqa
where in the last step we integrated by parts and used (\ref{source-f1}). This cancels the anomaly from (\ref{anomaly-pol-2d}).

\subsection{The 10d anomaly argument}

The anomalous Bianchi identity for $H_3$ is a crucial ingredient in
the Green-Schwarz anomaly cancellation mechanism in the critical 10d
heterotic theory. The extension (\ref{bianchi-new}) can therefore be
obtained by extending the anomaly cancellation analysis to
supercritical theory in the presence of a non-trivial normal
bundle. Using the fermion spectrum (\ref{fermions}), the anomaly
polynomial can be shown to factorize as $I_{12}=Y_4Y_8$, where $Y_8$
is a degree-8 polynomial in curvatures and field-strengths, and \beqa
Y_4=(\tr R^2-\tr F^2+\tr R_N^{\, 2}) \eeqa Incidentally, the
computation is isomorphic to that in
\cite{Schwarz:2001sf,Sugimoto:1999tx} in a different system
\footnote{Actually, this is because the fermion quantum numbers (\ref{fermions}) are identical to those arising in type I with additional D9-$\aDnine$ brane pairs \cite{Sugimoto:1999tx}, as noticed in \cite{Hellerman:2004zm}.}.  Anomaly cancellation is achieved through a
Green-Schwarz mechanism which involves a coupling $\int B_2 Y_8$ and
the Bianchi identity (\ref{bianchi-new}).

\subsection{Compactification and the 6d anomaly argument}

The integrated version of the Bianchi identity can be probed by
compactifying the 10d slice into a 4-manifold $X_4$, for instance K3
(as is familiar from supersymmetric compactifications). The K3
compactification of the gravitinos in (\ref{fermions}) produces an
anomaly exactly as in the critical case, see
e.g. \cite{Green:1984bx}. A direct test of eq. (\ref{bianchi-new}) is the computation of the pure gravitational anomaly, whose cancellation requires a net multiplicity of 244 positive chirality fermions in the theory. The spectrum of 6d massless fermions is determined by the index theorem in terms of characteristic classes of the tangent bundle, the gauge bundle, and the normal $SO(n)_{\rm rot}$ bundle.  

Consider the theory with $n=4$ supercritical dimensions, and introduce arbitrary instanton numbers $k,k'$ in $SU(2)\subset SO(4)$ factors of the gauge and normal bundles. The decomposition of the charged fermion spectrum in (\ref{fermions}) under the $SO(32)$ and instanton $SU(2)$ groups is
\beqa
& \quad\quad  SO(36)\times SO(4)_{\rm rot} & \supset SO(32)\times SU(2)\times SU(2)_{\rm rot}\nonumber \\
& {\rm Ferm}_+  \quad (\,\Yasymm,1) & \to (\,\Yasymm,1,1)+ 2(\fund,2,1) +\nonumber \\
&& \quad\quad + (1,3,1)+3(1,1,1)\nonumber \\
& {\rm Ferm}_- \quad (\fund,\fund) & \to \quad 2(\fund,1,2)+4(1,2,2)\nonumber \\
&{\rm Ferm}_+ \quad  (1,\Ysymm) & \to \quad 3(1,1,3)+(1,1,1)\nonumber
\eeqa
%
The net multiplicities we get for the different $SO(32)$ representations are:
\beqa
\# (\,\Yasymm\,)=-1 \;,\; \# (\fund)=k-k' \; ,\; \#({\bf 1})=-2(k-k')\quad\;\;
\eeqa
Cancellation of the gravitational anomaly requires that $k-k'=24$. This is the familiar statement that heterotic K3 compactifications requires instanton/NS5-brane number 24. In our setup, this number includes contributions from the extra gauge and normal bundles, which thus carry instanton/NS5-brane charge. Since the 10d gauge group $SO(32)$ is unbroken, configurations with different instanton numbers in the bundles $V$, $V'$ flow upon tachyon condensation to the critical heterotic theory with explicit NS5-branes (rather than finite size instantons).

\subsection{The worldsheet argument}
\label{throat}

Let us return to the case of flat non-compact 10d space, with $n=4$
supercritical dimensions, and non-trivial gauge and/or normal
bundles. We would like to analyze the worldsheet dynamics near the
vanishing locus of the (derivative of the) tachyon, which is the
Poincar\'e dual of the Euler class of the bundle. As above, we take
the instanton to be nontrivially fibered over a
$\bR^4\subset\bR^{9,1}$, parametrized by $X^\mu$ coordinates, and
take the gauge bundle to be nontrivial only for a $SO(4)\subset
SO(36)$ subgroup. The normal coordinates are still denoted by
$Y^m$. In the ansatz~(\ref{abs}) $\partial_n T^a$ vanishes at $X=0$,
so near the vanishing locus the $Y$ coordinates remain massless and
seemingly parametrize an extra branch of the CFT.

Integrating and requiring invariance under the  $\mathbb{Z}_2$ orbifold on the normal direction, the tachyon profile is locally of the form 
\beqa
T^a=\Gamma^\mu_{na} X^\mu Y^n\,.
\label{tach-zero}
\eeqa This conclusion is independent of which bundle has been twisted,
be it the normal or the gauge bundle. This tachyon profile couples to
the worldsheet fields through (\ref{tach-coupling}). In this case, the
only fields with non-trivial coupling to the tachyon profile are the
($4+4$) bosons $X^\mu$, $Y^m$, their right-moving fermion superpartners,
and the $4$ left-moving fermions $\lambda^a$ of $SO(4)\subset
SO(36)$. This matter content can be arranged in $(0,4)$ multiplets
(containing either four real right-moving fermions and four real
scalars, or one left-handed fermion).  The theory actually has $(0,4)$
susy at the level of interactions as well, since the Yukawa couplings
and the F-term potential turn out to obey the ADHM equations, as in
the analysis in \cite{ADHM} (not surprisingly since our system
describes instanton backgrounds). Moreover, the precise sigma model
arising from (\ref{tach-zero}) can be shown to correspond to the
singular CFT arising in the zero instanton size limit in
\cite{wshetinsts} (times the decoupled extra bosons and fermions,
which are not relevant for the discussion).

This can be shown as follows:  The scalar potential from (\ref{tach-zero}) is
 $\propto\vert X^2\vert\vert Y^2\vert$. If we split all
the $SO(4)$ indices into products of $SU(2)$'s as $a=YY'$, $m=AB$, $\mu=A'B'$
(with the appropriate reality conditions), one can recast \eqref{tach-zero}  as
$c^{CC'}\delta^Y_{A}\epsilon_{CB}\delta^{Y'}_{A'}\epsilon_{C'B'}X^{AY}Y^{A'Y'}$
for some appropriately normalized $c^{CC'}$ \footnote{This parameter
  is needed to translate from $(0,4)$ susy language to the $(0,1)$
  explicit from the tachyon. The fact that
  $C^{YY'}_{BB'}=\delta^Y_{A}\epsilon_{CB}\delta^{Y'}_{A'}\epsilon_{C'B'} X^{AY}Y^{A'Y'}$ satisfies the ADHM
  equations means that the resulting action is in fact independent of
  $c^{CC'}$.}. This is exactly the zero size instanton CFT in \cite{ADHM}. 

Thus, the infrared behavior of the system near the vanishing locus of the tachyon profile \eqref{tach-zero}  is exactly the CFT in a zero size instanton (equivalently,  NS5-brane) background. Deformations of this tachyon profile lead to marginal deformations in the IR CFT. As
argued in \cite{ADHM}, the Coulomb branch parametrized by the $Y$ is
decoupled from the physical (in our setup) Higgs branch parametrized
by the $X$ coordinates. Rather than physical spacetime coordinates
sticking out from $X=0$, the Coulomb branch signals that the
worldsheet theory is the singular CFT describing the presence of a
background NS5-brane (small instanton as in \cite{Witten:1995gx}) at
$X=0$.

\section{K-theory and other heterotic topological defects}

Closed string tachyon condensation allows for the
nucleation/annihilation of pairs of bundles associated to the
supercritical geometry and the gauge group. The corresponding (real)
K-theory thereby classifies the charges of topological defects
realized as closed string solitons. For instance, our construction of
the NS5-brane relates it to $KO(\IS^4)=\IZ$, and unveils a K-theoretic
structure in the anomalous Bianchi identity for $H_3$, refining its
earlier interpretation \cite{Wen:1985qj} in de Rahm or integer
cohomology --- analogous to the fact that RR fluxes in type II
theories are best understood in the context of K-theory
\cite{Moore:1999gb}.

The non-trivial real K-theory groups motivate the construction of other defects as closed tachyon solitons. In particular, the fundamental string can be identified as a soliton whose charge is $KO(\IS^8)=\IZ$, i.e. the second Pontryagin class at the cohomological level. This approach can also be exploited to construct torsion charges in the heterotic. For instance, the non-trivial element in $KO(\IS^2)=\IZ_2$ corresponds to a $\IZ_2$ charged 7-brane, built from tachyon condensation in $n=2$ supercritical dimensions, with normal or gauge $SO(2)\equiv U(1)$ bundles with
\beqa
\int_{\IR^2} F_{\rm gauge}-\int_{\IR^2} R_{\rm rot} =1\; {\rm mod}\; 2\IZ
\label{bundles-7brane}
\eeqa
This charge can be detected by considering its relative holonomy with the dual torsion object, a particle classified by $KO(\IS^9)=\IZ_2$. The latter is actually realized as a perturbative massive state, transforming as a bi-spinor of the $SO(32+2)$ gauge group and the normal $SO(2)$, which indeed picks up a $-1$ holonomy when moved around the defect associated to bundles satisfying (\ref{bundles-7brane}). It is also possible to show that the 7-brane $\IZ_2$ charge is detected as a global gauge anomaly \cite{Witten:1982fp} on an NS5-brane probe of the system, in the spirit of \cite{Uranga:2000xp}. The tachyon soliton realization of this 7-brane was proposed in \cite{Berasaluce-Gonzalez:2013sna}. Similarly, the groups $KO(\IS^1)=KO(\IS^{10})=\IZ_2$ are associated to closed tachyon solitons describing an 8-brane and an euclidean instanton.


\section{Conclusions}
In this letter we have obtained the heterotic NS5 brane as a soliton of the closed string tachyon in the supercritical heterotic string. The resulting supercritical soliton can be well understood using the perturbative worldsheet description of the supercritical string. We have checked the consistency of the construction using a variety of arguments both from the spacetime and worldsheet points of view.

Closed string tachyon condensation in the supercritical heterotic
provides a physical equivalence between non-trivial gauge fields and
spacetime geometry, in the sense that certain physical properties like
charges of topological defects, are preserved. This provides a
K-theoretic understanding of the heterotic anomalous Bianchi identity,
and thus of the involved charges. We expect that the generalization of these
ideas to other supercritical strings provides a new handle on the
non-perturbative branes in string theory. It is also tantalizing to
speculate that the equivalence we found between certain gauge and
spacetime topological configurations admits a dynamical extension into
physical processes translating non-trivial gauge dynamics into
modifications of the geometry, in the spirit of the phenomena found in
the holographic context.

\medskip

\begin{acknowledgments}
  \noindent{\bf Acknowledgments: } We thank A. Collinucci for illuminating
  discussions. MM and AU are partially supported by the grants  FPA2012-32828 from the MINECO, the ERC Advanced Grant SPLE under contract ERC-2012-ADG-20120216-320421 and the grant SEV-2012-0249 of the ``Centro de Excelencia Severo Ochoa" Programme. I.G.-E. would like to thank N.~Hasegawa for kind encouragement and support. MM is supported by a ``La Caixa'' Ph.D
  scholarship.

\end{acknowledgments}

\bibliography{refs}

\end{document}